\documentclass[aps,pra,preprint,showpacs,superscriptaddress]{revtex4}
\usepackage{amsmath,amssymb,latexsym}
\usepackage[T1]{fontenc}
\usepackage[english]{babel}
\usepackage{graphicx}

\begin{document}

\title{Separability conditions from the Landau-Pollak uncertainty relation}
\author{Julio I.\ \surname{de Vicente}} 
\affiliation{Departamento de Matem\'aticas, Universidad Carlos III
de Madrid, Avda.\ de la Universidad 30, 28911 Legan\'es, Madrid,
Spain}
\author{Jorge \surname{Sánchez-Ruiz}} 
\affiliation{Departamento de Matem\'aticas, Universidad Carlos III
de Madrid, Avda.\ de la Universidad 30, 28911 Legan\'es, Madrid,
Spain}
\affiliation{Instituto Carlos I de F\'{\i}sica Te\'orica y
Computacional, Universidad de Granada, 18071 Granada, Spain}

\begin{abstract}
We obtain a collection of necessary (sufficient) conditions for a bipartite system of
qubits to be separable (entangled), which are based on the Landau-Pollak formulation
of the uncertainty principle. These conditions are tested, and compared with
previously stated criteria, by applying them to states whose separability limits are
already known. Our results are also extended to multipartite and higher-dimensional
systems.
\end{abstract}


\maketitle

\section{Introduction}

Consider the vector $|\psi\rangle$, pertaining to a finite-dimensional Hilbert space
$H=H_A\otimes H_B$, that describes a pure state of two quantum systems $A$ and $B$.
$|\psi\rangle$ is said to be a product state if there exists $|\phi\rangle_A\in H_A$
and $|\varphi\rangle_B\in H_B$ such that
\begin{equation}\label{product}
|\psi\rangle=|\phi\rangle_A\otimes|\varphi\rangle_B \,.
\end{equation}
Separable states are mixtures of product states. In other words, the density operator
$\rho$ acting on $H$ that characterizes the quantum state of $A$ and $B$ is called
separable if it can be written as a convex combination of product vectors, that is,
\begin{equation}\label{separable}
\rho = \sum_ip_i|\phi_i,\varphi_i\rangle\langle\phi_i,\varphi_i| =
\sum_i p_i \, \rho^A_i\otimes\rho^B_i \,,
\end{equation}
where $0 \leq p_i \leq 1$, $\sum_i p_i = 1$, and
$|\phi_i,\varphi_i\rangle=|\phi_i\rangle_A\otimes|\varphi_i\rangle_B$.

If $\rho$ cannot be written as in Eq.\ (\ref{separable}), then the state is said to be
entangled. Entanglement is one of the most fascinating issues in quantum mechanics,
not only from a theoretical point of view \cite{epr}, but also because of its
applications in the context of quantum information theory, such as cryptography and
teleportation \cite{c&t}. Therefore, it is a very interesting question to ask whether
a given state is entangled or not. Although no general answer is known, there exist a
great variety of separability criteria, like the partial transpose criterion
\cite{parttrans}, Bell's inequalities violation \cite{bell}, and the construction of
entanglement witnesses (EW's) \cite{witness}. The first of these criteria gives
necessary and sufficient conditions when the dimension of $H$ is either $2\times2$ or
$2\times3$, while otherwise it is just a necessary condition. The second criterion
provides only a necessary condition. Finally, the third criterion is necessary and
sufficient in the sense that, given an entangled state, there always exists an EW that
detects it; however, it is not known how to construct all possible EW's, and this
criterion turns out to be a necessary separability condition once a particular set of
EW's has been chosen.

The relationship between entanglement and the uncertainty principle has been
investigated in several recent works (see e.g. \cite{duan}). The key fact is that,
when measuring a collection of nonlocal observables on a given state, the lower bound
on the uncertainty of the outcomes is higher for separable states than for entangled
states, because of the correlations inherent in the latter. Nonlocal operators
possess, in general, entangled eigenstates, while separable states cannot be
simultaneous eigenstates for the set of nonlocal operators. Using this idea, there
have been achieved variance-based separability criteria \cite{var} inspired by the
Heisenberg-Robertson formulation of the uncertainty principle \cite{hr}, as well as
entropy-based separability criteria \cite{gio,guh} derived from entropic uncertainty
relations \cite{eur,maa,san}. The necessary separability conditions obtained in this
way have the advantage of being more easily implemented in experiments, since they are
based on expectation values and probabilities for the outcomes of measurements. On the
contrary, the partial transpose criterion demands complete knowledge of the density
matrix, whose experimental determination requires considerable effort.

In this paper we derive new separability criteria based on a different mathematical
formulation of the uncertainty principle, the so-called Landau-Pollak uncertainty
relation, and we show that these conditions are better than those obtained using
entropies in the examples proposed so far. The article is organized as follows. The
Landau-Pollak uncertainty relation is briefly reviewed in Sec.\ II, where we state
some properties that will be useful later on. In Sec.\ III, we derive new separability
conditions for two-qubit systems. In Sec.\ IV, we investigate the accuracy of the
resulting criteria using some well-known examples. In Sec.\ V, the relationship
between one of our separability conditions and a set of optimal EW's is pointed out.
Section VI deals with the extension of our approach to more complex cases, i.e.\
bipartite systems of qudits and multipartite systems.

\section{The Landau-Pollak uncertainty relation}

Let $X$ denote a Hermitian operator representing some physical observable in a
finite-dimensional Hilbert space of dimension $D$, with a complete set of orthonormal
eigenvectors $\{|x_i\rangle\}$ $(i=1,2,\ldots,D)$ and $N$ distinct eigenvalues ($N\leq
D$). For $n=1,2,\ldots,N$, the probability $p_n(X)$ of finding the state $\rho$ in the
$n$th eigenspace of $X$ (i.e., the probability of obtaining the $n$th possible outcome
in a measurement of $X$) is given by
\begin{equation}\label{prob}
p_n(X) = \textrm{Tr} \big( P_n(X) \rho \big) \,,
\end{equation}
where $P_n(X)$ denotes the projection operator on the $n$th eigenspace of $X$.

The uncertainty principle states that, for general pairs of observables $X$ and $Y$,
the outcomes of a simultaneous measurement cannot both be fixed with arbitrary
precision. One way to express this fact mathematically is through the Landau-Pollak
uncertainty relation,
\begin{equation}\label{l-p}
\arccos\sqrt{\max_np_n(X)}+\arccos\sqrt{\max_np_n(Y)} \geq \arccos
c \,,
\end{equation}
where
\begin{equation}\label{c}
c = c(X,Y) \equiv \max_{i,j}|\langle x_i|y_j\rangle| \,.
\end{equation}
The relevance of this inequality in quantum mechanics was first pointed out by Uffink
\cite{uff}, who translated to the quantum language the original work of Landau and
Pollak on uncertainty in signal theory \cite{lan}.

The expressions
\begin{equation}\label{m}
M_r(\mathcal{P})=\left(\sum_{n=1}^N(p_n)^{1+r}\right)^{1/r}\,,
\quad r>-1 \,,
\end{equation}
measure the concentration of the probability distribution
$\mathcal{P}=(p_1,p_2,\ldots,p_N)$. They are closely related to the R\'enyi entropies
\cite{renyi},
\begin{equation}
H_q^{(R)}(\mathcal{P}) = \frac{1}{1-q} \ln \left( \sum_{n=1}^N (p_n)^q \right) \,,
\quad q>0 \,,
\end{equation}
and the Tsallis entropies \cite{tsallis},
\begin{equation}
H_q^{(T)}(\mathcal{P}) = \frac{1}{1-q} \left( \, \sum_{n=1}^N
(p_n)^q -1 \right) \,, \quad q>0 \,,
\end{equation}
both of which include the usual (Shannon) entropy as the particular case $q=1$. The
quantities $M_r(\mathcal{P})$ were first used as measures of uncertainty in quantum
mechanics in Refs.\ \cite{maa,uff}, where a summary of their properties is given; a
more detailed analysis can be found in \cite{hlp}. Here we will just mention that
$M_r(\mathcal{P})$ is a continuous non-decreasing function of $r$, with the limiting
value
\begin{equation}\label{limval}
M_\infty(\mathcal{P})=\max_{n}p_n \,,
\end{equation}
and $M_r(\mathcal{P})$ is convex in $\mathcal{P}$, i.e., for $0\leq\lambda\leq1$,
\begin{equation}\label{mrconvex}
M_r \big( \lambda\mathcal{P}_1+(1-\lambda)\mathcal{P}_2 \big) \leq
\lambda M_r(\mathcal{P}_1)+(1-\lambda)M_r(\mathcal{P}_2) \,.
\end{equation}

Taking into account Eq.\ (\ref{limval}), the Landau-Pollak uncertainty relation
(\ref{l-p}) can be written as
\begin{equation}\label{l-pm}
\arccos\sqrt{M_\infty(X)}+\arccos\sqrt{M_\infty(Y)}\geq\arccos c \,.
\end{equation}
Maximizing the sum $M_{\infty}(X)+M_{\infty}(Y)$ under the constraint (\ref{l-pm}), we
obtain the uncertainty inequality
\begin{equation}\label{l-pm-weak}
M_{\infty}(X)+M_{\infty}(Y) \leq 1+c \,,
\end{equation}
which is weaker than (\ref{l-pm}) but has a simpler and more natural form.

\section{Separability conditions for two-qubit systems}

Consider the following observables acting on a bipartite two-dimensional Hilbert
space,
\begin{equation}
Z=\sigma_z^A\otimes\sigma_z^B \,, \quad
X=\sigma_x^A\otimes\sigma_x^B \,,
\end{equation}
where $\sigma_i^j$ $(i=x,y,z;j=A,B)$ are the standard Pauli operators acting on the
$j$ qubit. Since $Z$ and $X$ commute, for this pair of observables we have that $c=1$,
and the right-hand side of (\ref{l-pm}) vanishes imposing no restriction on the
possible outcomes of measurements. The trivial lower bound 0 in Eq.\ (\ref{l-pm}) is
attained, for instance, if the measured state is one of the four maximally entangled
elements of the Bell basis,
\begin{align}\label{bbasis}
& |\phi^\pm\rangle = \frac{1}{\sqrt{2}}\big(|00\rangle\pm|11\rangle\big)\,, \nonumber \\
& |\psi^\pm\rangle =
\frac{1}{\sqrt{2}}\big(|01\rangle\pm|10\rangle\big)\,,
\end{align}
where we consider $|0\rangle$ and $|1\rangle$ to be eigenvectors of $\sigma_z$
corresponding to the eigenvalues $+1$ and $-1$, respectively.

However, if $Z$ and $X$ act on a separable state, the lower bound 0 is not attainable,
which enables the possibility of obtaining a separability condition. This can be done
by using Lemma 1 of \cite{guh}, which we quote here:

Let $\rho=\rho_A\otimes\rho_B$ be a product state on a bipartite Hilbert space
$H=H_A\otimes H_B$, and let $A$ ($B$) be observables with nonzero eigenvalues on $H_A$
($H_B$). Then
\begin{align} \label{lemma1}
& \mathcal{P}(A\otimes B,\rho)\prec\mathcal{P}(A,\rho_A), \nonumber \\
& \mathcal{P}(A\otimes B,\rho)\prec\mathcal{P}(B,\rho_B)
\end{align}
holds. The notation $\mathcal{P}\succ\mathcal{Q}$ (``$\mathcal{P}$ majorizes
$\mathcal{Q}$'') means that, if $\mathcal{P}=(p_1,p_2,\ldots,p_N)$ and
$\mathcal{Q}=(q_1,q_2,\ldots,q_N)$ denote two probability distributions written in
decreasing order (i.e. $p_1\geq p_2\geq\ldots\geq p_N$ and $q_1\geq q_2\geq\ldots\geq
q_N$), then
\begin{equation}\label{majorization}
\sum_{i=1}^{k}p_i\geq\sum_{i=1}^kq_i
\end{equation}
for all $k \in [1,\ldots,N]$.

It follows from the previous definition that Eq.\ (\ref{lemma1}) implies the
inequalities
\begin{align} \label{lemma1bis}
& M_\infty(A\otimes B,\rho)\leq M_\infty(A,\rho_A), \nonumber \\
& M_\infty(A\otimes B,\rho)\leq M_\infty(B,\rho_B).
\end{align}
Therefore, if $\rho_{sep}$ denotes an arbitrary (mixed) separable state, i.e.
$\rho_{sep}=\sum_ip_i\,\rho_i^A\otimes\rho_i^B$, and $A_1, A_2, B_1, B_2$ are
observables with nonzero eigenvalues, we have that
\begin{align}
& M_\infty(A_1\otimes B_1,\rho_{sep})+M_\infty(A_2\otimes
B_2,\rho_{sep})\,\nonumber\\
& \leq\sum_ip_i\big(M_\infty(A_1\otimes
B_1,\rho_i^A\otimes\rho_i^B)+M_\infty(A_2\otimes
B_2,\rho_i^A\otimes\rho_i^B)\big)\,\nonumber\\
& \leq \sum_i p_i \big( M_\infty(A_1,\rho_i^A)+M_\infty(A_2,\rho_i^A)
\big)\,\nonumber\\\label{inequality} &
\leq\sum_ip_i\big(1+c(A_1,A_2)\big)=1+c(A_1,A_2)\,,
\end{align}
where we have used Eqs.\ (\ref{mrconvex}) and (\ref{l-pm-weak}) in addition to
(\ref{lemma1bis}). Since both $\sigma_z$ and $\sigma_x$ have the eigenvalues $+1$ and
$-1$, they satisfy the conditions of the above lemma, and use of Eq.\
(\ref{inequality}) with the well-known value $c(\sigma_z,\sigma_x)=1/\sqrt{2}$ gives
\begin{equation}\label{separabilityweak}
M_\infty(Z,\rho_{sep})+M_\infty(X,\rho_{sep})\leq1+\frac{1}{\sqrt{2}}\approx1.71 \,.
\end{equation}

We have seen that the method developed by G\"{u}hne and Lewenstein in \cite{guh} to
derive separability conditions from entropic uncertainty relations can also be applied
to the Landau-Pollak uncertainty relation. However, as we shall prove in the
following, inequality (\ref{separabilityweak}) can be improved by performing a direct
maximization of the sum of $M_{\infty}(Z)$ and $M_{\infty}(X)$ in product states; the
bound attained in this way will be valid for any separable state because of the
convexity of $M_\infty$.

An arbitrary product state is of the form (\ref{product}) with
\begin{align} \label{pure}
& |\phi\rangle_A = \cos\alpha|0\rangle_A+e^{i\delta}\sin\alpha|1\rangle_A\,, \nonumber \\
& |\varphi\rangle_B =
\cos\beta|0\rangle_B+e^{i\gamma}\sin\beta|1\rangle_B\,,
\end{align}
where $\alpha,\beta\in[0,\pi/2]$ and $\delta,\gamma\in[0,2\pi)$. Both $Z$ and $X$ have
the eigenvalues $+1$ and $-1$, and the corresponding eigenspace projectors are
\begin{align}
& P_{+}(Z) = |00\rangle\langle00|+|11\rangle\langle11| \,, \nonumber \\
& P_{-}(Z) = |01\rangle\langle01|+|10\rangle\langle10| \,, \nonumber \\
& P_{\pm}(X)=|\phi^{\pm}\rangle\langle\phi^{\pm}|+|\psi^{\pm}\rangle\langle\psi^{\pm}|
\,.
\end{align}
Therefore, according to Eq.\ (\ref{prob}), the probabilities of finding the pure
separable state (\ref{product},\ref{pure}) in these eigenspaces are, respectively,
\begin{align}
& p_{+}(Z) = (\cos\alpha\cos\beta)^2+(\sin\alpha\sin\beta)^2 \,, \nonumber \\
& p_{-}(Z) = 1-(\cos\alpha\cos\beta)^2-(\sin\alpha\sin\beta)^2 \,, \nonumber \\
& p_{\pm}(X)=\frac{1}{2}(1\pm\cos\delta\cos\gamma\sin2\alpha\sin2\beta) \,.
\end{align}
Since $p_\pm(Z)$ do not depend on $\delta$ and $\gamma$, and $\sin2\alpha\sin2\beta$
is always nonnegative, the maximum value of $M_\infty(Z)+M_\infty(X)$ equals the
maximum of the functions
\begin{equation}\label{f}
f_{\pm}(\alpha,\beta)=p_\pm(Z)+\frac{1}{2}(1+\sin2\alpha\sin2\beta)\,,
\end{equation}
which occurs when $\alpha=\pm\beta$. Thus we find our first necessary separability
condition,
\begin{equation}\label{sep1}
M_\infty(Z,\rho_{sep})+M_\infty(X,\rho_{sep})\leq \frac{3}{2}\,.
\end{equation}
If for a certain state $M_\infty(Z)+M_\infty(X)>3/2$, then Eq.\ (\ref{sep1}) implies
that the state is entangled.

As shown in \cite{gio}, the introduction of a third observable,
\begin{equation}
Y=\sigma_y^A\otimes\sigma_y^B \,,
\end{equation}
enables the possibility of obtaining a more accurate separability condition, due to
the fact that we are then using the maximal number of complementary observables
available for each subsystem \cite{woo}. Unfortunately, no generalization of the
Landau-Pollak uncertainty relation is known for sets of more than two observables
(leaving aside the one that is trivially obtained from Eq.\ (\ref{l-pm-weak})), which
prevents us from using G\"{u}hne and Lewenstein's method in this case. Therefore, we
will follow the direct maximization procedure in order to set an upper bound for the
sum of $M_{\infty}(X)$, $M_{\infty}(Y)$, and $M_{\infty}(Z)$ in separable states.

Observable $Y$ has the same eigenvalues as $Z$ and $X$, with eigenspace projectors
\begin{equation}
P_\pm(Y)=|\phi^\mp\rangle\langle\phi^\mp|+|\psi^\pm\rangle\langle\psi^\pm|\,,
\end{equation}
and the corresponding probabilities for the pure separable state
(\ref{product},\ref{pure}) are
\begin{equation} \label{proby}
p_{\pm}(Y)=\frac{1}{2}(1\pm\sin\delta\sin\gamma\sin2\alpha\sin2\beta)\,.
\end{equation}
Since $\sin2\alpha\sin2\beta$ is nonnegative, and the maximum over $\delta$ and
$\gamma$ of the four functions of the form
$\pm(\sin\delta\sin\gamma\pm\cos\delta\cos\gamma) = \pm \cos (\delta \mp \gamma)$
equals 1, we only have to find the maximum of the functions
\begin{eqnarray}\label{g}
g_{\pm}(\alpha,\beta) & = & p_\pm(Z)+1+\frac{\sin2\alpha\sin2\beta}{2} \nonumber \\ &
= & f_{\pm}(\alpha,\beta) + \frac{1}{2} \,.
\end{eqnarray}
Recalling the derivation of Eq.\ (\ref{sep1}), we obtain our second necessary
separability condition,
\begin{equation}\label{sep2}
M_\infty(X,\rho_{sep})+M_\infty(Y,\rho_{sep})+M_\infty(Z,\rho_{sep})\leq2\,.
\end{equation}
Taking into account that $M_\infty(Y,\rho_{sep}) \geq 1/2$, we see that condition
(\ref{sep1}) can be derived from (\ref{sep2}), so that the latter is stronger than the
former.

Attending to \cite{var,gio}, the best separability conditions are obtained by choosing
as observables the three orthogonal components of the total spin of the system,
\begin{equation}\label{op3}
S_i=\sigma_i^A\otimes\mathbb{I}_B+\mathbb{I}_A\otimes\sigma_i^B\quad(i=x,y,z)\,,
\end{equation}
where $\mathbb{I}$ denotes the identity operator. These observables all have the
eigenvalues $\pm2$ (non-degenerate) and 0 (two-time degenerate), with eigenspace
projectors
\begin{align} \label{spinproj}
& P_{\pm}(S_x) =
  \frac{1}{2}\big(|\phi^+\rangle\pm|\psi^+\rangle\big)\big(\langle\phi^+|\pm\langle\psi^+|\big)
  \,, \nonumber \\
& P_{0}(S_x) = |\phi^-\rangle\langle\phi^-|+|\psi^-\rangle\langle\psi^-| \,, \nonumber \\
& P_{\pm}(S_y) =
  \frac{1}{2}\big(|\phi^-\rangle\pm|\psi^+\rangle\big)\big(\langle\phi^-|\pm\langle\psi^+|\big)
  \,, \nonumber \\
& P_{0}(S_y) = |\phi^+\rangle\langle\phi^+|+|\psi^-\rangle\langle\psi^-| \,, \nonumber \\
& P_{+}(S_z) = |00\rangle\langle00| \,, \quad P_{-}(S_z) = |11\rangle\langle11| \,, \nonumber \\
& P_{0}(S_z) = |01\rangle\langle01|+|10\rangle\langle10| \,,
\end{align}
and the corresponding probabilities for the generic pure state
(\ref{product},\ref{pure}) are
\begin{align} \label{prob3}
& p_{\pm}(S_x) =
  \frac{1}{4}(1\pm\cos\delta\sin2\alpha)(1\pm\cos\gamma\sin2\beta) \,, \nonumber \\
& p_{0}(S_x) = \frac{1}{2}(1-\cos\delta\cos\gamma\sin2\alpha\sin2\beta) \,, \nonumber \\
& p_{\pm}(S_y) =
  \frac{1}{4}(1\pm\sin\delta\sin2\alpha)(1\pm\sin\gamma\sin2\beta) \,, \nonumber \\
& p_{0}(S_y) = \frac{1}{2}(1-\sin\delta\sin\gamma\sin2\alpha\sin2\beta) \,, \nonumber \\
& p_{+}(S_z) = (\cos\alpha\cos\beta)^2 \,, \quad p_{-}(S_z) = (\sin\alpha\sin\beta)^2 \,,
  \nonumber \\
& p_{0}(S_z) = (\cos\alpha\sin\beta)^2+(\sin\alpha\cos\beta)^2 \,.
\end{align}
We therefrom see that the maximum value of $\sum_i M_\infty(S_i)$ for product states
is the maximum of
\begin{equation}
w(\alpha,\beta)=p(S_z)+1+\frac{\sin2\alpha\sin2\beta}{2}\,,
\end{equation}
which is easily found to be equal to 2. Thus we get our third necessary separability
condition,
\begin{equation}\label{sep3}
M_\infty(S_x,\rho_{sep})+M_\infty(S_y,\rho_{sep})+M_\infty(S_z,\rho_{sep})\leq2\,.
\end{equation}

Another interesting possibility is that of measuring a non-degenerate Bell diagonal
observable,
\begin{equation}\label{bobservable}
B = \lambda_1|\phi^+\rangle\langle\phi^+| + \lambda_2|\phi^-\rangle\langle\phi^-| +
\lambda_3|\psi^+\rangle\langle\psi^+| + \lambda_4|\psi^-\rangle\langle\psi^-| \,,
\end{equation}
with $\lambda_i\neq\lambda_j$ when $i\neq j$. The probability distribution for the
outcomes of $B$ acting on the pure separable state (\ref{product},\ref{pure}) is
\begin{align}
& p_{\phi^\pm}(B) = \frac{1}{2}
[(\cos\alpha\cos\beta)^2+(\sin\alpha\sin\beta)^2\pm\xi(\alpha,\beta)\zeta_+(\delta,\gamma)]\,,
\nonumber \\
& p_{\psi^\pm}(B) = \frac{1}{2}
[(\cos\alpha\sin\beta)^2+(\sin\alpha\cos\beta)^2\pm\xi(\alpha,\beta)\zeta_-(\delta,\gamma)]\,,
\end{align}
where $\xi(\alpha,\beta)=\frac{1}{2}\sin2\alpha\sin2\beta$ and
$\zeta_\pm(\delta,\gamma)=\cos(\delta\pm\gamma)$. The nonnegativity of
$\xi(\alpha,\beta)$ implies that $M_\infty(B)$ is the maximum over $\alpha$ and
$\beta$ of the functions
\begin{align}
& h_1(\alpha,\beta) =
\frac{1}{2}[(\cos\alpha\cos\beta)^2+(\sin\alpha\sin\beta)^2+\xi(\alpha,\beta)] \,,
\nonumber \\
& h_2(\alpha,\beta) =
\frac{1}{2}[(\cos\alpha\sin\beta)^2+(\sin\alpha\cos\beta)^2+\xi(\alpha,\beta)] \,,
\end{align}
and, therefore,
\begin{equation}\label{sep4}
M_\infty(B)\leq\frac{1}{2}\,.
\end{equation}
This last necessary separability condition is not new, since it was previously derived
by G\"{u}hne and Lewenstein \cite{guh} using a different method. As pointed out by
these authors, condition (\ref{sep4}) is equivalent to the set of four optimal EW's
\begin{equation}\label{ew}
W_{\phi^\pm}=\frac{1}{2}\mathbb{I}-|\phi^\pm\rangle\langle\phi^\pm|\,,\quad
W_{\psi^\pm}=\frac{1}{2}\mathbb{I}-|\psi^\pm\rangle\langle\psi^\pm|\,.
\end{equation}

\section{Accuracy of the separability conditions}

Next we will test the power as entanglement detectors of the separability conditions
derived in the previous section, by applying them to states whose separability limits
are already known. We will also compare our separability conditions with previous
criteria. All the probabilities below are calculated using Eq.\ (\ref{prob}) and the
projectors found in Sec.\ III.

\subsection{Werner states}

Werner states \cite{wer} are mixtures of a completely random state and a maximally
entangled pure state. In the case of two qubits, and choosing the maximally entangled
state to be the singlet state, they read
\begin{equation}\label{werner}
\rho_W=\frac{1-p}{4}\mathbb{I}_A\otimes\mathbb{I}_B+p|\psi^-\rangle\langle\psi^-|\,,
\end{equation}
where $p\in[0,1]$. These states are known to be separable iff $p\leq1/3$ (see
\cite{pit} and references therein). The probabilities of finding $\rho_W$ in each
eigenspace when measuring the observables of Sec.\ III are
\begin{align}\nonumber
& p_\pm(X)=p_\pm(Y)=p_\pm(Z)=\frac{1\mp p}{2}\,,\\\nonumber &
p_0(S_i)=\frac{1+p}{2}\,,\\\nonumber
& p_\pm(S_i)=p_{\phi^\pm}(B)=p_{\psi^+}(B)=\frac{1-p}{4}\,,\\
& p_{\psi^-}(B)=\frac{1+3p}{4}\,.
\end{align}
Thus we have,
\begin{align}\nonumber
& \sum_{\tau=X,Z}M_\infty(\tau,\rho_W)=1+p\,,\\\nonumber &
\sum_{\tau=X,Y,Z}M_\infty(\tau,\rho_W)=\frac{3(1+p)}{2}\,,\\\nonumber
& \sum_{i=x,y,z}M_\infty(S_i,\rho_{W})=\frac{3(1+p)}{2}\,,\\
& M_\infty(B,\rho_W)=\frac{1+3p}{4}\,.
\end{align}

We see from these results that the separability condition (\ref{sep1}) detects
entanglement when $p>1/2$, while (\ref{sep2}), (\ref{sep3}), and (\ref{sep4}) detect
entanglement when $p>1/3$. It is worth noting that in this case the three latter
separability conditions, like variance-based criteria \cite{var}, are optimal in the
sense that they are able to detect all the entangled states. All four conditions
improve the bound obtained in \cite{gio} using Shannon entropies ($p>0.55$), as well
as those derived in \cite{guh} by means of Tsallis entropies ($p>1/\sqrt{3}$) and
Bell's inequality criterion ($p>1/\sqrt{2}$). Even more, when measuring the same
observables (i.e.\ when using the same experimental setting), our conditions always
improve on the bounds given by the Shannon and Tsallis entropic conditions,
respectively: $p>0.78$ and $p>1/\sqrt{2}$ when measuring $X$ and $Z$; $p>0.65$ and
$p>1/\sqrt{3}$ when measuring $X$, $Y$, and $Z$; $p>0.55$ when measuring $S_x$, $S_y$,
and $S_z$; and $p>0.74$ when measuring $B$ (in the last two cases only Shannon
entropic conditions are available).

\subsection{Gisin states}

Gisin states \cite{gis} are mixtures of the same fraction of the pure states
$|00\rangle$ and $|11\rangle$, and any pure superposition of the states $|01\rangle$
and $|10\rangle$. That is,
\begin{equation}\label{gisin}
\rho_G=p|\chi\rangle\langle\chi|+\frac{1-p}{2} \big(
|00\rangle\langle00|+|11\rangle\langle11| \big) \,,
\end{equation}
where $|\chi\rangle=\cos\alpha|01\rangle+e^{i\beta}\sin\alpha|10\rangle$,
$\alpha\in[0,\pi/2]$, $\beta\in[0,2\pi)$, and $p\in[0,1]$. The state $\rho_G$ is known
to be separable iff \cite{parttrans}
\begin{equation}
p\leq\frac{1}{1+\sin2\alpha}\,.
\end{equation}
In this case,
\begin{align}\nonumber
& p_{\pm}(X) = p_{\pm}(Y) = \frac{1 \pm p\sin2\alpha\cos\beta}{2} \,,\\\nonumber &
p_+(Z)=1-p,\quad p_-(Z)=p\,,\\\nonumber &
p_\pm(S_x)=p_\pm(S_y)=\frac{1}{2}p_+(X)\,,\\\nonumber &
p_0(S_x)=p_0(S_y)=p_-(X)\,,\\\nonumber
& p_\pm(S_z)=p_{\phi^\pm}(B)=\frac{1-p}{2},\quad p_0(S_z)=p\,,\\
& p_{\psi^\pm}(B)=\frac{p \, (1 \pm \sin2\alpha\cos\beta)}{2}\,,
\end{align}
which leads to
\begin{align}
& \sum_{\tau=X,Z}M_\infty(\tau,\rho_G)=\max\{p,1-p\} + \frac{1 + p\sin2\alpha \left|
\cos\beta \right|}{2}\,, \nonumber \\
& \sum_{\tau=X,Y,Z}M_\infty(\tau,\rho_G)=\max\{p,1-p\}+1+p\sin2\alpha
\left| \cos\beta \right|\,, \nonumber \\
& \sum_{i=x,y,z}M_\infty(S_i,\rho_{G})=\max\left\{p,\frac{1-p}{2}\right\} \nonumber \\
& + 2\max\left\{\frac{1-p\sin2\alpha\cos\beta}{2},
\frac{1+p\sin2\alpha\cos\beta}{4}\right\}\,, \nonumber \\
& M_\infty(B) = \max \left\{ \frac{1-p}{2}, \frac{p \, (1+\sin2\alpha \left| \cos\beta
\right|)}{2}\right\} \,.
\end{align}

These results imply that conditions (\ref{sep1}), (\ref{sep2}), (\ref{sep3}), and
(\ref{sep4}) detect entanglement when $p > \big( 1+\frac{1}{2}\sin2\alpha \left|
\cos\beta \right| \big)^{-1}$, $p>(1+\sin2\alpha \left| \cos\beta \right|)^{-1}$,
$p>(1-\sin2\alpha\cos\beta)^{-1}$, and $p>(1+\sin2\alpha \left| \cos\beta
\right|)^{-1}$, respectively (notice that the restriction imposed by (\ref{sep3}) is
meaningful only when $\beta\in(\pi/2,3\pi/2)$). Thus we find that in this case the
best separability conditions are (\ref{sep2}) and (\ref{sep4}), though in general they
are not optimal. When $\beta = 0, \pi$ all entangled states are detected by
(\ref{sep2}) and (\ref{sep4}), but as $\beta$ departs from these values the
separability conditions fail to detect an increasing amount of entangled states, until
for $\beta = \pi/2, 3\pi/2$ no entanglement is detected. For values of $\beta$ such
that $\left| \cos \beta \right| > \sqrt{2}-1$, conditions (\ref{sep2}) and
(\ref{sep4}) improve the bound given by Bell's inequality criterion, $p > \big(
1+(\sqrt{2}-1)\sin2\alpha \big)^{-1}$ \cite{gis}. It is worth noting that, due to the
dependence of the probabilities on two parameters, to establish which states are
detected by the entropic separability criteria is mathematically cumbersome, and has
to be carried out by numerical analysis.

\subsection{Mixtures of a singlet and a maximally polarized pair}

The states
\begin{equation}
\rho_0=p|\psi^-\rangle\langle\psi^-|+(1-p)|00\rangle\langle00|\,,
\end{equation}
with $p \in [0,1]$, are known by the positive partial transpose criterion to be
separable only if $p=0$ \cite{parttrans}. The probabilities for the observables of
Sec.\ III are now
\begin{align}\nonumber
& p_\pm(X)=p_\pm(Y)=\frac{1\mp p}{2}\,,\\\nonumber &
p_0(S_x)=p_0(S_y)=\frac{1+p}{2}\,,\\\nonumber &
p_\pm(S_x)=p_\pm(S_y)=\frac{1-p}{4}\,,\\\nonumber &
p_+(Z)=p_+(S_z)=1-p\,,\\\nonumber &
p_-(Z)=p_0(S_z)=p_{\psi^-}(B)=p\,,\\\nonumber &
p_-(S_z)=p_{\psi^+}(B)=0\,, \\ & p_{\phi^\pm}(B)=\frac{1-p}{2}\,,
\end{align}
and, therefore,
\begin{align}\nonumber
&
\sum_{\tau=X,Z}M_\infty(\tau,\rho_0)=\max\{p,1-p\}+\frac{1+p}{2}\,,\\\nonumber
&
\sum_{\tau=X,Y,Z}M_\infty(\tau,\rho_0)=\max\{p,1-p\}+1+p\,,\\\nonumber
& \sum_{i=x,y,z}M_\infty(S_i,\rho_{0})=\max\{p,1-p\}+1+p\,,\\
& M_\infty(B) = \max \left\{ p,\frac{1-p}{2} \right\} \,.
\end{align}

We thus find that condition (\ref{sep1}) detects entanglement for $p>2/3$, while it
suffices to have $p>1/2$ in order to detect entangled states using conditions
(\ref{sep2}), (\ref{sep3}), and (\ref{sep4}). These bounds are not optimal, but they
improve on that derived from the violation of Bell's inequality, $p>0.8$
\cite{parttrans}. Furthermore, as in the case of Werner states, in each measurement
setting the bounds provided by our conditions are better than those obtained using
Shannon entropies: $p>0.85$ when measuring $X$ and $Z$; $p>0.73$ when measuring $X$,
$Y$, and $Z$; $p>0.55$ when measuring $S_x$, $S_y$, and $S_z$; and $p>0.78$ when
measuring $B$.

\section{Equivalence of condition (\ref{sep2}) and the set of optimal EW's (\ref{ew})}

The necessary separability condition (\ref{sep4}) is equivalent to the set of optimal
EW's (\ref{ew}) \cite{guh}. Using the three-dimensional space representation of
density matrices with coordinates Tr$(X\rho)$, Tr$(Y\rho)$, and Tr$(Z\rho)$ (see
Refs.\ \cite{guh} and \cite{3d}), this equivalence means that condition (\ref{sep4})
is able to recognize the octahedron containing all separable states, which lies inside
the tetrahedron whose vertices are the Bell states and contains all possible states.

For the three families of states considered in the previous section, the separability
conditions (\ref{sep2}) and (\ref{sep4}) detect the same entangled states, which
suggests that they are equivalent. In the following we will prove that this is indeed
the case, so that condition (\ref{sep2}) is also equivalent to the set of optimal EW's
(\ref{ew}) and has the same success at detecting the octahedron that contains the
separable states.

Condition (\ref{sep4}) can be stated as
\begin{equation} \label{sep4-alt}
0 \leq \textrm{Tr} \big( |BS_i\rangle\langle BS_i|\rho_{sep} \big) \leq \frac{1}{2}\,,
\end{equation}
where $|BS_i\rangle$ is any element of the Bell basis (\ref{bbasis}). Taking into
account the identities \cite{guh}
\begin{align} \label{qaz}
& \textrm{Tr} \big( |\phi^{\pm}\rangle\langle\phi^{\pm}|\rho \big) =
\frac{1{\pm}\textrm{Tr}(X\rho){\mp}\textrm{Tr}(Y\rho)+\textrm{Tr}(Z\rho)}{4}\,,
\nonumber \\
& \textrm{Tr} \big( |\psi^{\pm}\rangle\langle\psi^{\pm}|\rho \big) =
\frac{1{\pm}\textrm{Tr}(X\rho){\pm}\textrm{Tr}(Y\rho)-\textrm{Tr}(Z\rho)}{4}\,,
\end{align}
and noting that
\begin{equation}
\textrm{Tr}(\tau\rho) = p_+(\tau)-p_-(\tau) \equiv \Delta p(\tau) \quad (\tau=X,Y,Z)
\,,
\end{equation}
the inequalities in (\ref{sep4-alt}) can be written as
\begin{align}
& -1 \leq {\pm} \Delta p(X,\rho_{sep}) {\mp} \Delta p(Y,\rho_{sep}) + \Delta
p(Z,\rho_{sep}) \leq 1 \,, \nonumber \\
& -1 \leq {\pm} \Delta p(X,\rho_{sep}) {\pm} \Delta p(Y,\rho_{sep}) - \Delta
p(Z,\rho_{sep}) \leq1 \,.
\end{align}
This is equivalent to the eight inequalities of the form
\begin{equation}
-1\leq\pm\Delta p(X,\rho_{sep})\pm\Delta p(Y,\rho_{sep})\pm\Delta
p(Z,\rho_{sep})\leq1\,,
\end{equation}
that is,
\begin{equation}\label{qwerty}
|\Delta p(X,\rho_{sep})|+|\Delta p(Y,\rho_{sep})|+|\Delta p(Z,\rho_{sep})|\leq1\,.
\end{equation}
Finally, noting that for $\tau=X,Y,Z$
\begin{equation}
|\Delta p(\tau)| = M_\infty (\tau) - \big( 1-M_\infty (\tau) \big)
= 2M_\infty (\tau)-1\,,
\end{equation}
Eq.\ (\ref{qwerty}) reduces to (\ref{sep2}), which proves that this separability
condition is equivalent to (\ref{sep4}).

\section{Separability conditions for more complex systems}

If we consider multipartite and/or higher-dimensional systems (qudits), the direct
maximization procedure used in Sec.\ III for two-qubit systems becomes too complicated
to be carried out analytically, due to the increasing number of free parameters,
although it can be faced numerically. However, the method of G\"{u}hne and Lewenstein
(see Sec. III) can also be applied in this case, and allows us to derive separability
conditions from the Landau-Pollak uncertainty relation.

\subsection{Bipartite systems of qudits}

For states of a two-dimensional Hilbert space, the best detection of entanglement is
achieved by measuring in each subsystem the three orthogonal components of spin, which
are also a maximal set of complementary observables. We recall that two observables
$A,B$ in $D$-dimensional Hilbert space are said to be complementary if
$c(A,B)=1/\sqrt{D}$ \cite{sch}, and maximal sets of $D+1$ pairwise complementary
observables are known to exist when $D$ is either a prime \cite{iva} or a power of a
prime \cite{woo}. However, when the dimension of the Hilbert space is greater than
two, the orthogonal components of spin are not complementary observables and both
cases must be treated separately.

Choosing $A_1$, $A_2$ and/or $B_1$, $B_2$ to be complementary observables in
$D$-dimensional Hilbert space, we find from Eq. (\ref{inequality}) that
\begin{equation}\label{quditcompl}
M_\infty(A_1\otimes B_1,\rho_{sep})+M_\infty(A_2\otimes
B_2,\rho_{sep})\leq1+\frac{1}{\sqrt{D}} \,.
\end{equation}

On the other hand, if $S_{n}$ and $S_{n'}$ denote $D$-dimensional spin observables
along the axes $n$ and $n'$, respectively, we have that \cite{san2}
\begin{align}\label{san2}
& c^2(S_{n},S_{n'})=\left(
\begin{array}{c}
  D-1 \\
  n^\ast \\
\end{array}
\right) \left(\cos^2\frac{\beta}{2}\right)^{D-1-n^\ast}
\left(\sin^2\frac{\beta}{2}\right)^{n^\ast}, \nonumber \\
& n^\ast = \left[ D\sin^2 \frac{\beta}{2} \right],
\end{align}
where $\beta$ is the angle between the axes $n$ and $n'$, and the square brackets
denote integer part of the expression within. Therefore, use of Eq.\
(\ref{inequality}) leads to
\begin{align}
& M_\infty(S_n^A\otimes S_{n}^B,\rho_{sep})+M_\infty(S_{n'}^A\otimes
S_{n'}^B,\rho_{sep})\,\nonumber\\\label{quditspin} & \leq1+\sqrt{\left(
\begin{array}{c}
  D-1 \\
  n^\ast \\
\end{array}
\right)}\left(\cos\frac{\beta}{2}\right)^{D-1-n^\ast}\left(\sin\frac{\beta}{2}\right)^{n^\ast},
\end{align}
and choosing the axes $n, n'$ to be orthogonal ($\beta = \pi /2$)
the previous inequality simplifies to
\begin{align}
& M_\infty(S_x^A\otimes S_{x}^B,\rho_{sep})+M_\infty(S_{z}^A\otimes
S_{z}^B,\rho_{sep})\,\nonumber\\\label{quditspinort} & \leq1+\sqrt{\frac{1}{2^{D-1}}
\left(
\begin{array}{c}
  D-1 \\
  \left[ D/2 \right] \\
\end{array}
\right)} \;.
\end{align}

It is worth noting that when $D$ is odd the spin observables have one non-degenerate
zero eigenvalue, so that the conditions in G\"{u}hne and Lewenstein's lemma are not
fulfilled. However, as pointed out by these authors \cite{guh}, the requirement that
the observables have nonzero eigenvalues is more a technical condition and can always
be achieved by altering the eigenvalues, since the Landau-Pollak uncertainty relation,
like the entropic uncertainty relations considered in \cite{guh}, does not depend on
them.

\subsection{Multipartite systems}

In the case of tripartite systems we must distinguish between fully separable states,
which are states (or mixtures of states) of the form
\begin{equation}\label{fullysep}
|\psi\rangle_{ABC}=|\phi\rangle_A\otimes|\varphi\rangle_B\otimes|\chi\rangle_C \,,
\end{equation}
and biseparable states, which are product states with respect to one particular
bipartite splitting of the system, e.g.
\begin{equation}\label{bisep}
|\psi\rangle_{ABC}=|\phi\rangle_A\otimes|\varphi\rangle_{BC} \,,
\end{equation}
or mixtures of states of this form. Fully separable and biseparable states, as well as
other kinds of partially separable states, can be defined likewise for general
multipartite systems.

A straightforward generalization of Eq.\ (\ref{inequality}) enables us to derive
biseparability conditions for multipartite qubit and qudit systems. Thus, for
instance, on the analogy of (\ref{separabilityweak}) we find the following
biseparability condition for systems of three qubits:
\begin{align}\label{bi}
& M_\infty(\sigma_x^A\otimes\sigma_x^B\otimes\sigma_x^C) +
M_\infty(\sigma_z^A\otimes\sigma_z^B\otimes\sigma_z^C) \nonumber \\
& \leq1+\frac{1}{\sqrt{2}}\approx1.71 \,.
\end{align}
Likewise, the multipartite analogues of Eqs.\ (\ref{quditcompl}) and
(\ref{quditspinort}) are, respectively, the following biseparability conditions for
systems with an arbitrary number of subsystems in $D$-dimensional Hilbert space:
\begin{align}
& M_\infty(C_1^{A_1}\otimes\cdots\otimes
C_1^{A_D},\rho_{sep})+M_\infty(C_{2}^{A_1}\otimes\cdots\otimes
C_{2}^{A_D},\rho_{sep})\,\nonumber\\
& \leq1+\frac{1}{\sqrt{D}}\,,
\end{align}
where $C_1$ and $C_2$ are complementary observables, and
\begin{align}
& M_\infty(S_x^{A_1}\otimes\cdots\otimes
S_x^{A_D},\rho_{sep})+M_\infty(S_{z}^{A_1}\otimes\cdots\otimes
S_{z}^{A_D},\rho_{sep})\,\nonumber\\
& \leq1+\sqrt{\frac{1}{2^{D-1}}\left(
\begin{array}{c}
  D-1 \\
  \left[ D/2 \right] \\
\end{array}
\right)} \;.
\end{align}

We emphasize that, as already noted in Sec. III in relation to the two-qubit case, the
separability and biseparability conditions obtained in this section cannot be improved
by considering measurements of additional observables, due to the fact that no
nontrivial generalization of the Landau-Pollak uncertainty relation is known for sets
of more than two observables \cite{nota}.

\section{Conclusions}

We have derived several necessary separability conditions for two-qubit systems,
namely Eqs.\ (\ref{sep1}), (\ref{sep2}), (\ref{sep3}), and (\ref{sep4}), on the basis
of the so-called Landau-Pollak uncertainty relation. Like entropy-based separability
criteria, our conditions are expressed in terms of the probability distributions for
the outcomes of measurements, so that they can be applied in many experimental
settings. On the other hand, the measure of uncertainty used here, $M_\infty$, is
mathematically easier to handle than entropies.

In order to test the power of these conditions as entanglement detectors, we have
applied them to three well-known families of two-qubit states, namely Werner states,
Gisin states, and mixtures of a singlet and a maximally polarized pair. In most cases,
the results obtained are better than those provided by other separability criteria,
such as Bell's inequalities violation and entropy-based criteria. Conditions
(\ref{sep2}), (\ref{sep3}), and (\ref{sep4}) are even able to detect all entangled
two-qubit Werner states, thus improving on entropy-based criteria \cite{gio,guh} and
reproducing the results of variance-based criteria \cite{var}. However, the other two
families show that in general our conditions are not optimal, i.e.\ they are necessary
but not sufficient. It would be interesting to know whether a refined choice of
operators can give optimal results for these states, and, more generally, whether
given an entangled state it is always possible to construct a set of observables such
that the sum of their $M_\infty$ measures is greater in that state than in a generic
product state.

We have proved that conditions (\ref{sep2}) and (\ref{sep4}) are equivalent. Since
(\ref{sep4}) is known to be equivalent to the set of four optimal EW's (\ref{ew}), the
same happens for (\ref{sep2}). As a consequence, (\ref{sep2}) is able to detect all
entangled states lying outside the octahedron of separable states in the
three-dimensional representation of density matrices \cite{nonlinearEW}. Condition
(\ref{sep1}) is weaker than (\ref{sep2}), since it does not include the correlations
in the third observable; however, we have considered it explicitly because it only
needs two measurements and, therefore, it is experimentally less demanding.

Finally, we have extended our results to more complicated cases than two-qubit
systems, i.e. to multipartite and higher-dimensional systems, for which no necessary
and sufficient condition for entanglement is known to date. The separability
conditions obtained in these cases, however, are limited due to the lack of a
nontrivial uncertainty relation of Landau-Pollak type for sets of more than two
observables. Therefore, further research in this field might help to improve the
results presented here.

\begin{acknowledgments}
The work of the second author (J.S.R.) has been partially supported by Dirección
General de Investigación (Ministerio de Ciencia y Tecnología) of Spain under grants
BFM2001-3878-C02-01 and BFM 2003-06335-C03-02, and the Junta de Andalucía research
group FQM-0207.
\end{acknowledgments}


\begin{thebibliography}{widest-label}
\bibitem{nota} On the contrary, nontrivial entropic uncertainty relations are known
for maximal sets of complementary observables in arbitrary dimension \cite{san}.
However, in Ref.\ \cite{guh} the authors do not seem to take advantage of this fact
for the case of tripartite systems of qubits, since they just consider a
three-observable inequality that can be deduced trivially from a two-observable one.
It is also worth noting that spin observables might have been used as well to derive
separability conditions for qudit systems from entropic uncertainty relations.
\end{thebibliography}
\end{document}